\documentclass[12pt,letterpaper]{JHEP3}
\usepackage{cite}
\usepackage{epsfig}
\usepackage{amsfonts}

\renewcommand{\thanks}[1]{\footnote{#1}} % Use this for footnotes

\newcommand{\be}{\begin{equation}}
\newcommand{\bea}{\begin{eqnarray}}
\newcommand{\eea}{\end{eqnarray}}
\newcommand{\beq}{\begin{equation}}
\newcommand{\ee}{\end{equation}}
\newcommand{\eeq}{\end{equation}}
\def\ba{\begin{eqnarray}}
\def\ea{\end{eqnarray}}
\def\bd{\begin{displaymath}}
\def\ed{\end{displaymath}}

\def\O{{\cal O}}
\def\N{{\cal N}}

\title{Chronology Protection in String Theory}

\author{
Lisa Dyson\\
Center for Theoretical Physics\\
Massachusetts Institute of Technology\\
Cambridge, MA  02139\\

Institute for Theoretical Physics\\
Stanford University\\
Stanford, CA 94305\\

E-mail: \email{ldyson@mit.edu}}

\abstract{
Many solutions of General Relativity appear to allow the possibility 
of time travel.  This was initially a fascinating discovery, but 
geometries of this type violate causality, a basic physical law 
which is believed to be fundamental.  Although string theory is 
a proposed fundamental theory of quantum gravity, geometries with 
closed timelike curves have resurfaced as solutions to its low 
energy equations of motion.  In this paper, we will study the 
class of solutions to low energy effective supergravity theories 
related to the BMPV black hole and the rotating Wave--D1--D5--brane 
system.   Time travel appears to be possible in these geometries.  
We will attempt to build the causality violating regions and propose 
that stringy effects prohibit their construction.  We will show
how the geometry is corrected and that, once corrected, causality is 
preserved. We will track our chronology protection proposal in the
dual conformal field theory.   The absence of
closed timelike curves in the geometry coincides with the 
preservation of unitarity in the conformal field theory.
The agent of chronology protection for the geometries studied here
mirrors the enhancon mechanism, a mechanism string theory employs to 
resolve a class of 
naked singularities.}
 
\preprint{\hepth{0302052} \\ MIT-CTP 3360 \\ SU-ITP 03-09}

\begin{document}

%%%%%%%%%%%%%%%%%%%%%%%%%%%%%%%%%%%%%%%%%%%%%%%%%%%%%%%%%%%%%%%%%%%%%%%%%%%%%%%
\section{Introduction}
General Relativity is believed to be an accurate description of our classical world.  While this led to a revolution in physical thought in the twentieth century, it also introduced a myriad of ambiguities and physical conundrums.  One such ambiguity arises when attempting to understand the quantum description of black holes.  Naive calculations seem to indicate that information may be lost and thereby destroy our notion of unitary time evolution \cite{loss}.  Physicists have battled over this issue for decades.  As a proposed ultimate theory of quantum gravity, string theory has intervened and provided some answers.  For a certain class of black hole geometries, a microscopic description exists in string theory which enables one to calculate macroscopic quantities such as entropy and Hawking radiation \cite{entropy, non}.  Resulting arguments seem to indicate that there is no loss of information and the unitarity of physics is preserved \cite{su, witten, eternal}.

Another class of solutions to General Relativity which have caused much debate are geometries that are not black holes, but nevertheless have regions of infinite curvature, that is, geometries with singularities un-shielded by horizons.  Such geometries are ubiquitous in General Relativity, but believed to be physically unsound. Again, string theory has intervened and provided insight.  There exist mechanisms in string theory which resolve naked singularities, resulting in consistent, non-singular physics \cite{ks, ps, enhancon}.

General relativity also admits geometries that have closed timelike curves.  Examples include the Godel universe \cite{godel}, Kerr black holes \cite{kerr, carter} and Gott's time machines \cite{gott}.  Naively this would indicate that time travel is possible.  This in turn gives birth to a whole host of physical inconsistencies such as the Grandmother paradox, where one is allowed to modify the initial conditions that lead to ones own existence, and the Bootstrap paradox, where an effect is its own cause \footnote{See, for example, \cite{cpa2} for a review}.

There has been much work trying to address the paradoxes posed by these geometries.  Some have accepted the possibility of time travel, while others have argued that physical processes come into play rendering time travel impossible.  In the case of Gott's time machine, for instance, it has been shown that any attempt to create a time machine in a closed universe would cause the space to collapse entirely \cite{gott1, gott2, gott3}.  Many believe that quantum physics will ultimately prevent all geometries which naively allow time travel from forming.  This led to a proposal by Stephen Hawking who stated  ``It seems that there is a Chronology Protection Agency which prevents the appearance of closed timelike curves and so makes the universe safe for historians.'' \cite{cpa}

As of yet, the full consistency of geometries of this type has not been tested in string theory \footnote{See \cite{horava} for a recent discussion on holography in supersymmetric Godel universes.}.  

In this work, we will propose that a Chronology Protection Agency does indeed exist in string theory.  Specifically, we will study solutions to ten dimensional supergravity which can be dimensionally reduced to give five dimensional rotating black holes.  When the rotation parameter exceeds a certain value, closed timelike curves appear outside of the horizon.  By studying the full ten dimensional solution, we are able to track down the source of causality violations.  We will find that our solution breaks down long before we reach the causally sick region.  We conclude that the geometry must be altered at this point.
The causality violating region is never created and our usual notion of chronology is retained.

In the next section, we will explore the geometry in detail, first presenting the five dimensional black hole solution, then its ten dimensional counterpart.  We will review how closed timelike curves form in both the five dimensional and ten dimensional geometries and how the causality bound is associated with a unitarity bound in the dual conformal field theory.  This bound, however, does not prohibit the causally undesirable geometry from forming.  In fact, supersymmetric solutions appear to exist for arbitrary values of angular momentum.

In order to determine if there is some form of an obstruction to creating the causality violating region which would have a non-unitary conformal field theory dual, in section three, we examine the objects associated with the charges that make up the geometry.  We propose that there is indeed an obstruction to forming the geometry and that chronology is protected.  The agency of protection for the class of geometries that we discuss is analogous to the enhancon mechanism presented in \cite{enhancon} which resolves a class of naked singularities. 
In section four we study the conformal field theory dual in the context of the AdS/CFT duality.  Unitarity requires that the naive solution be corrected.  We show that unitarity is indeed preserved once the geometry has been corrected.  We conclude in section five.

\section{The Geometry}

\subsection{Five Dimensional Black Hole}

Let us begin by presenting the five dimensional rotating black hole solution.  We will consider a black hole with three charges, call them $Q_1, Q_5$, and $Q_k$.  The metric for this geometry is

\begin{eqnarray}
ds^2  &=&  - (f_1 f_5 f_k)^{-{2 \over 3}}\Big[ dt^2 + {J \over 2 r^2} (\sin^2 \theta d \phi_1 - \cos^2 \theta d \phi_2)\Big]^2 \nonumber \\ 
 && \quad \quad \quad \quad + (f_1 f_5 f_k)^{1 \over 3} \Big[ dr^2 + r^2 d \theta^2 + r^2 (\sin^2 \theta d \phi_1^2 + \cos^2 \theta d \phi_2^2) \Big]
\end{eqnarray}
where, as usual, the $f_i$ are harmonic functions associated with the charges, $f_i = 1 + Q_i/r^2$.  In addition to the metric, this supergravity solution has the following moduli and gauge fields under which the black hole is charged,

\be
e^{-2\phi} = {f_5 \over f_1} \quad \quad \quad e^{2 \sigma} = { f_k \over f_5^{1/4} f^{3/4}} \quad \quad\quad e^{\sigma_i} = {f_1^{1/4} \over f_5^{1/4}} 
\label{bhmoduli}
\ee
\be
A =  {1 \over f_1} \ \Big[- {Q_1 \over r^2} dt + { J\over 2 r^2} (\sin^2 \theta d \phi_1 - \cos^2 \theta d \phi_2)\ \Big] 
\ee
\be
A^k =  {1\over f_k} \ \Big[-{Q_k \over r^2} dt + { J\over 2 r^2} (\sin^2 \theta d \phi_1 - \cos^2 \theta d \phi_2)\ \Big]
\ee
\be
B =  Q_5 \cos^2\theta d\phi_1 \wedge d \phi_2 +  {1 \over f_1} { J\over 2 r^2 } dt \wedge (\sin^2 \theta d \phi_1 - \cos^2 \theta d \phi_2)
\label{blackhole}
\ee
with $i=1,...4$.

This geometry has been studied extensively by many authors \cite{bmpv, cvetic, myersbh, herdeiro2, herdeiro}.  In order to have a supersymmetric rotating black hole, the horizon must be stationary.  In \cite{myersbh} it was shown that the horizon is not rotating.  The nonzero rotation parameter affects the horizon by turning it into a squashed sphere.  Since the squashing parameter depends on the angular momentum, $J$, one can show that a sensible description of the horizon can breakdown if $J$ becomes too large.  

It was additionally shown that this geometry has closed timelike curves.  Luckily, if $J$ is small enough, all closed timelike curves are hidden behind the horizon.  However, when $J$ becomes too large, closed timelike curves appear outside of the horizon leading to gross causality violations.  The full sickness of this over-rotating black hole geometry was studied in detail in \cite{herdeiro2}.  

The causality constraint on $J$ can be seen explicitly by looking at the angular components of the metric.  The chronology horizon, $R_{ch}$, is the location where $g_{\phi_i \phi_i}$ vanishes.  Closed timelike curves appear when $g_{\phi_i \phi_i}$ becomes negative.  This occurs if $J^2 > 4 \,(r^2 + Q_1) (r^2 + Q_5) (r^2 + Q_k)$.  Since the horizon is located at the origin in these coordinates, an observer outside of the horizon will see closed timelike curves if $J^2 > 4\, Q_1Q_5Q_k$.  

This constraint on $J$ also becomes apparent when studying the thermodynamics of the black hole.  Its Bekenstein-Hawking entropy is

\be
S = { \pi^2 \over 2 G_5} \sqrt{ Q_1 Q_5 Q_k - { J^2 \over 4}}
\label{entropy}
\ee
This can become imaginary if $J$ is too large.  To be complete, let us also specify the mass of the black hole

\be
M = { \pi\over 4 G_5} ( Q_1 + Q_5 + Q_k)
\label{mass}
\ee

We can attempt to understand the origin of the causally sick region by lifting this five dimensional solution to ten dimensions and studying the fundamental objects which create the black hole.  We will begin this process by presenting the ten dimensional geometry below.

\subsection{Ten Dimensional D--Brane Geometry}

Upon lifting to ten dimensions, we find a supergravity description of D1 and D5--branes with momentum running along the effective D--string \cite{herdeiro}.  The D5--branes are additionally wrapped on $T_4$.  The metric, RR field and the dilaton are 

\begin{eqnarray}
ds^2 &=& {1 \over \sqrt{f_1 f_5}} \Big[ -dt^2 + {Q_k \over r^2} (dz - dt)^2 + dz^2 + { J \over r^2} (\sin^2 \theta d\phi_1 - \cos^2\theta d\phi_2)(dz - dt)\Big] \nonumber\\
& & + \sqrt{{f_1 \over f_5}}\, ds^2_{T^4} +  \sqrt{f_1 f_5}\, \Big[ dr^2 + r^2 d \theta^2 + r^2 (\sin^2 \theta d \phi_1^2 + \cos^2 \theta d \phi_2^2)\Big]
\label{fullgeo}
\end{eqnarray}
\be
C^{(2)} = {1 \over f_1} dt \wedge dz +  {1 \over f_1}{J \over 2 r^2 } (\sin^2 \theta d\phi_1 - \cos^2\theta d\phi_2) \wedge (dz - dt) + Q_5 \ \cos^2 \theta d \phi_1 \wedge d\phi_2
\label{rr}
\ee
\be
e^{-2 \Phi} = {f_5 \over f_1}
\ee
The charges $Q_1, Q_5$, $Q_k$ and $J$ are given by

\be
Q_1 = { g \alpha'^3 \over V} N_1, \quad \quad   Q_5 = g \alpha' N_5, \quad   \quad Q_k = { g^2 \alpha'^4 \over R^2 V} N_k, \quad \quad J = { g^2 \alpha'^4 \over R V} F_R
\label{charges}
\ee
where we have $N_1$ D1--branes, $N_5$ D5--branes, $N_k$ units of right moving momentum, and the angular momentum is quantized in terms of integers $F_R$.  $V$ is the volume of $T^4$ and $R$ is the radius of the $S^1$.
  
This is a IIB supergravity solution with four supercharges.  We can consider curves which have tangent vectors given by  
\be
l^{\mu} \partial_{\mu} = \alpha \, \partial _{z} + \beta \,(\partial_{\phi_1} -  \partial_{\phi_2})\ 
\label{cyclech}
\ee
\cite{herdeiro}.
Since $z$ is a compact direction, for certain values of $\alpha$ and $\beta$, curves of this type can be closed.  Also, a quick calculation of the proper length shows that these curves can be timelike in the region $r < R_{ch}$. Closed timelike curves reappear in ten dimensional language.
We have not been able to successfully rid ourselves of the sickness associated with the five dimensional geometry 

In fact, we can do better than this.  We can T-dualize our solution to find a system which should have the same physics.  In the new geometry, the full sickness of this configuration becomes apparent.  Performing a T-duality along the $z$ direction, we find a solution to IIA supergravity with the following metric.

\bea
ds^2 &=& {1\over f_k \sqrt{f_1 f_5}} \ \Big[ -dt^2 + {J \over 2 r^2} ( - \sin^2\theta d\phi_1+ \cos^2\theta d\phi_2) \ dt \Big] + \sqrt{ f_1 \over f_5} \ ds^2_{T^4} 
 \nonumber \\
&& + \, \sqrt{f_1 f_5} \Bigg[ \left( 1 - { J^2 \over 4} {\sin^2 \theta \over r^6 f_1 f_5 f_k} \right) r^2 \sin^2\theta \, d\phi_1^2 + \left( 1 - { J^2 \over 4} {\cos^2\theta \over r^6 f_1 f_5 f_k}  \right) r^2 \cos^2\theta \, d\phi_2^2 \Bigg] \nonumber \\
&& +
\,  { J^2 \over 2} {1 \over r^4 f_k \sqrt{f_1 f_5}} \,\cos^2 \theta  \, \sin^2\theta \, d\phi_1 \, d\phi_2\ + \sqrt{f_1 f_5} \, \Big[ {dz^2\over f_k} + dr^2 + r^2 d\theta^2\Big]
\label{tdual}
\eea
\\
Since $g_{\phi_i \phi_i}$ can become negative, we see clearly that for large enough $J$ causality violations are an integral part of ten-dimensional physics.  The causally sick region in the higher dimensional geometry coincides precisely with the causally sick region of the lower dimensional black hole.

\subsection{The Dual Field Theory} 
\label{CFT}

The geometry is not the only place where we find a breakdown in our physical description of this solution.  Recall that there is a gauge theory that describes this configuration \cite{entropy, bmpv}.  If we take the size of the $S^1$ to be much larger than the size of the $T^4$, the effective description is the 1+1 dimensional  field theory living on the world volume of the D1--branes.  In the decoupling limit, we have the near horizon limit of our D--brane geometry, AdS$_3 \times S^3 \times T^4$, which is dual to an $\N$ = 4 conformal field theory living on the boundary of the AdS$_3$ .

Turning off angular momentum for a moment, the supersymmetric states of the field theory that we are interested in are in the left moving ground state with arbitrary right moving charge.  Since these states are BPS, the energy and momentum are related by $E \sim N_k/R$, where $R$ is the radius of the $S^1$.  

Adding rotation requires that we use up some of the energy of the oscillator modes.  
We can break up the $N=4$ superconformal algebra into left and right moving $N=2$ algebras.  There is a $U(1)$ subgroup associated with each and we can specify the charges as $(F_R, F_L)$. 

One can bosonise the $U(1)$ currents \cite{bmpv} to show that a state with charge $F_R$ is represented by an operator of the form 
\be
\O = \exp \left({i \sqrt{3} F_R\phi \over \sqrt{c}}\right) \Phi
\label{state}
\ee
where $\Phi$ is an operator from the rest of the field theory.   
The eigenvalue of $L_0$ gives the conformal weight of primary operators.  Since $L_0(\O) = N_k$,
\be
L_0(\Phi) = N_k  - {F_R^2 \over 4 N_1 N_5} 
\label{cftlevel}
\ee
Unitarity requires that conformal weights be positive.  We see then that $N_k \ge F_R^2/(4N_1N_5)$, or equivalently, $Q_k \ge J^2/(4Q_1Q_5)$.  This is nothing more than the causality bound that we saw previously.  So we see that the unitarity bound of the conformal field theory coincides with the causality bound of the geometry \cite{herdeiro}.  Let us also note that (\ref{cftlevel}) represents the effective level of the system after we have used some energy to add angular momentum.  The level determines the degeneracy of states.  The entropy can be easily calculated and agrees with the Bekenstein- Hawking entropy (\ref{entropy}) calculated above  \cite{bmpv}.

We have seen that the physics of this solution breaks down in three dual descriptions: the five dimensional black hole, the ten dimensional D--brane configuration and the two dimensional conformal field theory.  In an attempt to understand how we might resolve the causally sick region, we will study the  ten dimensional supergravity solution a bit further below.

\section{The Resolution}

\subsection{The Geometry}
We can imagine building our geometry by slowly bringing in D--brane charge from infinity. A probe calculation will determine if this is a consistent thing to do.  
The action for a D1--brane is

\be
S = -\tau_1 \int d^2 \sigma \, e^{-\Phi} \sqrt{-det \, g} + \tau_1 \int C^{(2)}
\ee
Working in static gauge $ t = \tau , z = \sigma$, and choosing the ansatz $ x^i = x^i(\tau), y^a = y^a$,
where the $x^i$ run over the transverse directions and the $y^a$ run over the $T^4$, the slow velocity limit of the action is
\be
S =  \tau_1 \int d^2 \sigma  \ f_1 f_5 f_k  \ \Big[{\dot r}^2 + r^2 ({\dot \theta}^2 + \sin^2 \theta {\dot \phi_1}^2 + \cos^2 \theta {\dot \phi_2}^2)\Big]  
\label{action}
\ee
In this limit, the potential vanishes leaving behind a purely kinetic term in the action.  We can therefore consistently bring D1--branes to the origin.  A similar analysis holds for D5--branes (after dualizing the RR 2-form to get the appropriate RR 6-form under which the D5--branes are electrically charged).  So there is no obstacle here.  Where, then, is the problem? 

We have probed this geometry with objects associated with two of the charges that make up the geometry.  Recall that there is also charge associated with momentum modes running along the effective D--string, $Q_k$.  This is nothing more than a gravitational wave.  In a T-dual picture, this charge is associated with a fundamental string.  There is additional momentum, $J$, associated with rotation.  In order to determine where the breakdown in our geometry may be occurring, let us consider its microscopic description.

We can think of a bound state of one D1--brane and $N_5$ D5--branes as a system of $N_5$ fractional strings \cite{mathur1,fatblack, mathur2}.  If we have $N_1$ D1--branes, there are a total of $N_1 N_5$ fractional strings.  In this model, angular momentum can be added to the system by allowing each sub-string to be in a particular spin state of the $SU_R(2) \times SU_L(2)$ symmetry \cite{mathur2}.  We can join together any number of the sub-strings.  The spin of the string created, however, must be the same as that of a single string.  The maximum angular momentum of the system, then, occurs when all of the strings are split and aligned along the same direction.  The angular momentum of this state is ( ${N_1 N_5 \over 2}, {N_1 N_5 \over 2} $).

There are also massless open strings in our system which begin and end on the D--branes.  When we add $N_k$ units of momentum along these massless strings, the  entropically favored configuration is one in which all of the fractional strings join and form a single long string with an effective length of order $R' =N_1 N_5 R$, where $R$ is the radius of the $S^1$ along which the D-branes are wrapped \cite{fatblack}.  The effective level of the string state becomes $N' = N_1 N_5 N_k$.   If we want to add angular momentum to this system, it must be carried by the massless strings since the spin of the single long string formed by the D--branes is effectively zero \cite{mathur2}.  The effective level, once we add angular momentum, becomes 

\be
N' = N_1 N_5 \left( N_k  - {F_R^2 \over 4 N_1 N_5} \right).
\label{level}
\ee
The standard thermodynamic quantities can easily be calculated in this model.  Since open strings are attached to the horizon in this limit, they account for the degrees of freedom which give rise to the entropy of the black hole.
The entropy is $S = 2 \pi \sqrt{N'}$.   This agrees with (\ref{entropy}).

Returning to the supergravity picture, we want to form the three charge black hole.  We therefore assume that the D--branes have fractionalized and joined together to form one long string.  We have seen from the above probe calculation (\ref{action}) that we can consistently bring D1 and D5--branes in from infinity to the origin to form our geometry.    What about the momentum modes with non-zero rotation which, in the supergravity description, correspond to rotating gravitational waves?

We will attempt to add rotating gravitational waves to a D1-D5 system by doing the following.  Assume we have an interior geometry made up of only D1 and D5--brane charge.  The metric for this system is:

\be
ds^2 = {1 \over \sqrt{f_1 f_5}} \Big[ -dt'^2 + dz'^2 \Big] + \sqrt{{f_1 \over f_5}}\, ds^2_{T^4} +  \sqrt{f_1 f_5}\, \Big[ dr^2 + r^2 d \theta^2 + r^2 (\sin^2 \theta d \phi_1'^2 + \cos^2 \theta d \phi_2'^2)\Big]
\label{d1d5}
\ee
We will join this to the geometry (\ref{fullgeo}) which additionally has rotating gravitational waves.  We will join these two geometries together at some radius $R$.  Continuity of the metric requires the coordinates to be related by the following transformations:

\bea
\phi_1' & = & \phi_1 + {J \over 2 R^4 F_1 F_5} \ (z - t) \nonumber \\
\phi_2' & = & \phi_2 - {J \over 2 R^4 F_1 F_5} \ (z - t) \nonumber \\
z' & = & z + {1 \over 2 R^2}\left( Q_k - { J^2 \over 4 R^4 F_1 F_5}\right) (z - t) \nonumber \\
t' & = & t + {1 \over 2 R^2} \left( Q_k  - { J^2 \over 4 R^4 F_1 F_5}\right) (z - t)
\label{coords}
\eea
where $F_i = (1 + Q_i/R^2)$.

Our new geometry has D1 and D5--branes at the origin and rotating gravitational waves localized on the shell at radius $R$.  The standard Israel matching conditions \cite{israel} give the tension and pressures associated with the joining shell.  With this, we can determine if it is consistent to bring the gravitational waves in to the origin and hence create the geometry which is described by equation (\ref{fullgeo}) for all $r$.  The energy-momentum tensor for the shell is 
$S^{\mu}_{\nu} = 8 \pi G_{10} \sqrt{G_{rr}} \ T^{\mu}_{\nu}$ with

\be
T_0^0 = - T_z^z = T_0^z =  -T_z^0 = {2 \over R^3} \Bigg(Q_k - {J^2 \over 4 R^4 F_1 F_5}  \Bigg)
\label{tensor}
\ee
\bd 
T_0^{\phi_1} =  - T_0^{\phi_2} =  T_z^{\phi_2} = - T_z^{\phi_1} = {J \over  R^5 F_1^2 F_5^2} \ g(R) \nonumber 
\ed
\bd
T_{\phi_1}^0 = T_{\phi_1}^z = -  { J\over  R^3 F_1 F_5} \ g(R) \   \sin^2 \theta 
\ed
\bd
T_{\phi_2}^0 = T_{\phi_2}^z =   {J \over R^3 F_1 F_5} \ g(R) \ \cos^2 \theta \nonumber 
\ed
\bd
T_{\alpha,\beta} = 0  \nonumber 
\ed
where ${\alpha,\beta}$ runs over the transverse directions and the $T^4$, $G_{10} =  8 \pi \alpha'^4 g^2$ is the ten dimensional Newton's constant and $g(R) = [ 2 + (R^2 F_1 F_5)'/(R F_1 F_5)]$.   The energy-momentum tensor is all we need determine where the breakdown of our geometry occurs.  But let us stop here for a moment to confirm that it is indeed what we expect.

First, notice that the stresses and energy vanish in the transverse direction.  This tells us that there are no transverse forces acting on the shell to prohibit it from forming and it is consistent to bring it in from infinity.
Next,  for large $R$, the asymptotic tension of the shell, $T_0^0$, goes like the inverse of the area of a three sphere.  This gives precisely the charge-energy relationship that we expect for a gravitational wave, $E \sim Q_k$, and it contributes the appropriate amount to the asymptotic mass of the black hole (\ref{mass}).  Finally, the non-vanishing off-diagonal angular components of the stress energy tensor encode the rotation of the shell.

The tension of the shell,  
\be
\tau ={2 \over R^3} \Bigg(Q_k - {J^2 \over 4 R^4 F_1 F_5} \Bigg)
\label{fulltension}
\ee
represents the local energy density of the incoming gravitational waves from the supergravity perspective and massless strings from the microscopic D--brane point of view.  Placing the shell at a radius $R$, the effective local Kaluza Klein momentum $E \sim P'$ is
\be
P' \sim Q_k - {J^2 \over 4 R^4 F_1 F_5}
\label{momentum}
\ee
Recall that momentum corresponds to the mass of Kaluza Klein particles in a compactified theory.  Hence, we have Kaluza Klein particles with $m \sim P^\prime$.   

If $J> 4 \, Q_1 Q_5 Q_k$, the tension and momentum of the shell vanishes before the chronology horizon is reached.  This occurs at a radius

\be
R_{cp}^2 = - {(Q_1 + Q_5) \over 2} + {1 \over 2} \sqrt{(Q_1 - Q_5)^2 + {J^2 \over  Q_k}}. 
\label{cp}
\ee
If we place the shell at this radius, the energy density of our constructed geometry is smooth and the mass of the Kaluza Klein particles vanishes.  New massless states arise in the compactified theory which imply that the low energy theory must be corrected.  If we try to bring the shell of gravitational waves in past $R_{cp}$, the energy density of the shell becomes negative.  In order to avoid negative energy densities, we conclude that the gravitational waves are unable to travel past $R_{cp}$.  This occurs precisely at the point in moduli space where we expect corrections to the supergravity theory due to the presence of the massless Kaluza Klein particles.  Since the causally sick region is never created, chronology is protected.

Let us stop for a moment to recap.  The naive supergravity solution is made up of D1--branes, D5--branes, and massless strings carrying momentum which are represented by gravitational waves.  Following \cite{mathur2, fatblack}, we have argued that the rotation in our geometry is carried by the gravitational waves.  We were able to bring D1 and D5--branes in to the origin.  When we tried to bring in rotating gravitational waves from infinity to complete the construction of our geometry, we found that negative energy densities appear before the chronology horizon is reached.  In order to avoid negative energy densities, we propose that the gravitational waves are not able to travel beyond a chronology protection radius, $R_{cp}$.  This radius is the location in moduli space where Kaluza Klein particles become massless.  It is also the location where the energy of the shell due to momentum along the compact direction exactly cancels the energy due to rotation.  The ten dimensional geometry, therefore, has a smooth energy density. The final corrected geometry has momentum charge localized on the shell at radius $R_{cp}$ outside of the would-be chronology horizon with D1 and D5--branes at the origin.  The potential causality violating region is never able to form and chronology is protected.

We should note that although we wrapped the D5--branes on $T^4$, we could have wrapped them on K3.  Depending on the values of the parameters, wrapping branes on K3 can have the undesirable effect of creating naked singularities of repulson type.  Luckily, these singularities can easily be resolved as shown in \cite{enhanconbh, enhanconrot}.  The extension of our result to this case is straight forward.  Since the origin of the causality violating region are the rotating gravitational waves and not the D1 or D5--branes, 
%the resolution of the repulson singularity proceeds as in \cite{enhanconbh, enhanconrot}.  In order to prevent any causality violations that might occur for large angular momentum, a shell of rotating gravitational waves forms at the radius $R_{cp}$.    
the resulting geometry will have a shell of D1 and D5--branes placed at the appropriate radius to resolve the repulson singularity and a shell of rotating gravitational waves placed at the chronology protection radius $R_{cp}$, as discussed here, to prevent closed timelike curves from forming.

\subsection{Repulsive Coulomb Force}
Let us return to the five dimensional geometry (\ref{bhmoduli}) - (\ref{blackhole}).  The ten dimensional rotating gravitational wave becomes a particle charged under the gauge field $A_k$ in five dimensions.  We can probe our geometry with a point particle with charge $q$ which couples to the gauge field $A_k$.  The Hamiltonian for such a particle is \cite{herdeiro}.
\begin{eqnarray}
H &=& g^{\mu\nu} ( p_{\mu} + q A^k_{\mu})(p_{\nu} + q A^k_{\nu})
\end{eqnarray}

Let us focus on the Coulomb potential that the particle feels moving radially in this background. 
\be
V = q \ g^{\mu\nu} ( p_{\mu} A^k_{\nu} + p_{\nu} A^k_{\mu}) 
 = {q Q_k \over r^2} \left(1 - {J^2 \over 4 Q_k  r^4 f_1 f_5} \right) \sqrt{f_1 f_5}
\ee
The Coulomb interaction changes sign precisely at $r=R_{cp}$, if the Coulomb force was attractive for $r>R_{cp}$, it becomes negative for $r<R_{cp}$ and visa versa \cite{herdeiro}.  So it appears as though there is some non-trivial charged object localized at the surface $r=R_{cp}$.  Our resolved geometry simply places it there.  Again, this is analogous to the mechanism of \cite{enhancon, consistency} where a point particle probe interacting with the gravitational field felt nontrivial mass localized on a surface.  The resolved geometry in that case simply placed the appropriate massive objects (D6-D2*--branes) at that location.  

In fact, as we saw previously, the asymptotic tension of the shell goes like $ Q_k/ A_3$.  This is precisely the relationship that we expect for a particle charged under the gauge field $A_k$ with its charge smeared over a three sphere.  Recall also that the shell is located at a safe distance away from the causally sick region, $R_{cp}> R_{ch}$.   So the end result is that a rotating shell with asymptotic charge $Q_k$ and local charge density given by (\ref{fulltension}) forms outside of the chronology horizon and thereby prevents any attempt to create closed timelike curves.  The interior region is no longer sick.  It is just a standard compactified D1--D5--brane geometry.

Up to this point, we have only used the microscopic description of our solution to determine which objects carry the charges.  A similar analysis was carried out for geometries with naked singularities of repulson type in \cite{consistency}.  There is a general rule here.   The limits of physical consistency are exceeded before the sick regions can be created.  For a large class of geometries which are physically unsound, a simple geometric study of the laws of physics alone can reveal how one might correct a geometry to render it physically sound.

\subsection{Massless Strings}

Returning now to the microscopic description of our configuration,
recall that the naive geometry is a solution to Type IIB effective low energy supergravity with momentum modes and no winding modes.  In a T-dual picture, momentum modes and winding modes are interchanged.  A T-dual string, then, has winding number and hence mass proportional to (\ref{momentum}).  
%If we try to bring a string in beyond $R_{cp}$, its mass would be negative.  This would be unphysical. 
 If we try to construct the T-dual geometry, as we bring the strings in from infinity to create the geometry, the mass of the strings vanish at the chronology protection radius, $R_{cp}$.  When massive modes become light, their contribution to the low energy effective theory can no longer be ignored.  The low energy theory must be corrected at this point to include the new massless degrees of freedom.  

Happily, massless modes appear at the chronology protection radius, $R_{cp}$.  The breakdown in the supergravity theory coincides precisely with the breakdown we encountered when trying to construct the geometry above, since we encountered negative energy densities beyond $R_{cp}$. Once the new massless degrees of freedom have been taken into account, we propose that the above geometry is, to leading order, the solution to the corrected  supergravity theory.  
%This also applies in the limit that the radius of the $S^1$ goes to infinity.  Hence, the closed timelike curves that appear in the compact theory are a result of compactifying a region that needs to be corrected.  

\section{Unitarity in the Conformal Field Theory}
We have proposed that the naive BMPV black hole and D1-D5-Wave geometries cannot be constructed when the angular momentum parameter exceeds a certain bound, $J^2 > 4 Q_1 Q_5 Q_k$.  Instead, a shell carrying $Q_k$ and $J$ charge forms outside of the would be causally sick region.  The charges needed to create the causality violating region are not permitted to travel past this shell.  Closed timelike curves, therefore, are not able to form and the resulting geometry is causally safe.  

We can take the decoupling limit of the repaired geometry to study the AdS space which is dual to a conformal field theory.  As we saw previously, the causality bound of our spacetime coincides with a unitarity bound in the conformal field theory \cite{herdeiro}.  We can take the limit of our resolved geometry to see if it agrees with the expected result.

Assuming that the angular momentum exceeds the unitarity bound, the shell of gravitational waves forms outside of the horizon at the chronology protection radius, $R_{cp}$.  Zooming in on the near horizon region gives the near horizon geometry of the D1--D5--brane configuration of (\ref{d1d5}).  This is nothing more than the zero mass BTZ black hole.  The unitarity violations discussed in section \ref{CFT} have been remedied.  The dual conformal field theory of the corrected geometry is unitary.

\section{Conclusion}
As a fundamental theory of quantum gravity, string theory has been shown to successfully resolve many of the paradoxes inherent in Einstein's theory of General Relativity.  Examples include non-singular descriptions of geometries with regions of infinite curvature and evidence that information is not lost in black hole geometries.  We propose that string theory also has something to say about closed timelike curves.

We studied the BMPV rotating black hole and its higher dimensional Wave-D1-D5--brane analogue.  Naively, closed timelike curves appear outside of the horizon if the rotation parameter exceeds a certain value.  When trying to construct the geometry, however, we saw that massless degrees of freedom arise before reaching the causally sick region.  These new degrees of freedom appear at precisely the location where the energy due to transverse momentum cancels the energy due to rotation of the gravitational waves which make up our geometry.  We propose that the special radius where this occurs is a chronology protection horizon, beyond which the gravitational waves cannot travel.  Instead they expand to form a shell outside of the chronology violating region.  Since the presence of the gravitational waves beyond this point was crucial to create the causally sick region, our geometry is rendered safe and chronology is protected.  

If $J$ exceeds the causality bound in the naive solution, the dual conformal field theory would be non-unitary.  The correction to the geometry that has been proposed is not only causally safe, but has a unitarity conformal field theory dual.

The agent of chronology protection for the class of geometries studied here has a natural physical interpretation.  
Assume we begin with the standard low energy solution (\ref{fullgeo}) with $J < 4 \, Q_1 Q_5 Q_k$.  If we increase the angular momentum so that this bound is no longer satisfied, the gravitons in the ten dimensional language and the Kaluza Klein particles in the lower dimensional language begin to expand.  This expansion is due to the centrifugal repulsion and causes the gravitons (particles) to grow to just the right size to compensate for the repulsive force acting on them.  The end result is that giant gravitons (particles) form well outside of the chronology horizon to prevent the causally sick region from being created

It is important to note that the configurations studied in this paper are a part of a larger class of geometries related by a series of dualities.  For instance, the dyonic black hole and NS5--NS1--wave configurations considered in \cite{tseytlin} are S-dual to the backgrounds studied here.  The chronology protection arguments should have a dual description in these geometries.

Time travel gives rise to a host of physical inconsistencies.  One may conclude that any fundamental theory of quantum gravity would not admit solutions where time travel is possible.  The low energy effective equations of string theory have as solutions an abundance of such causality violating geometries \cite{horava,cks,supertubes,5dsolns}.  We believe, however, that string theory is causally sound.
As we have argued for the class of geometries studied here, 
stringy corrections to the low energy effective theory resolve causality violations.  We propose that once all stringy effects are taken into account,
 our usual notion of chronology will emerge as a protected law of nature.

\section{Acknowledgments}

I would like to thank B. Freivogel, S. Ganguli, S. Hellerman, P. Horava, S. Kachru, M. Kleban, S. Shenker, E. Silverstein, L. Susskind, and U. Varadarajan for useful discussions.

\end{document}